\documentclass{aa}
\usepackage{graphics,epsfig,amsmath,amssymb,amstext,txfonts}
\def\la{\mathrel{\mathchoice {\vcenter{\offinterlineskip\halign{\hfil
$\displaystyle##$\hfil\cr<\cr\sim\cr}}}
{\vcenter{\offinterlineskip\halign{\hfil$\textstyle##$\hfil\cr
<\cr\sim\cr}}}
{\vcenter{\offinterlineskip\halign{\hfil$\scriptstyle##$\hfil\cr
<\cr\sim\cr}}}
{\vcenter{\offinterlineskip\halign{\hfil$\scriptscriptstyle##$\hfil\cr
<\cr\sim\cr}}}}}

\def\ga{\mathrel{\mathchoice {\vcenter{\offinterlineskip\halign{\hfil
$\displaystyle##$\hfil\cr>\cr\sim\cr}}}
{\vcenter{\offinterlineskip\halign{\hfil$\textstyle##$\hfil\cr
>\cr\sim\cr}}}
{\vcenter{\offinterlineskip\halign{\hfil$\scriptstyle##$\hfil\cr
>\cr\sim\cr}}}
{\vcenter{\offinterlineskip\halign{\hfil$\scriptscriptstyle##$\hfil\cr
>\cr\sim\cr}}}}}

\begin{document}

\title{On the viability of holistic cosmic-ray source models}

\author{J. Aublin \and E. Parizot}

\institute{Institut de Physique Nucl\'eaire d'Orsay, IN2P3-CNRS/Universit\'e Paris-Sud, 91406 Orsay Cedex, France}

\offprints{parizot@ipno.in2p3.fr}

\date{Received date; accepted date}

\abstract{We consider the energy spectrum of cosmic-rays (CRs) from a purely phenomenological point of view and investigate the possibility that they all be produced by the same type of sources with a single power-law spectrum, in $E^{-x}$, from thermal to ultra-high energies. We show that the relative fluxes of the Galactic (GCR) and extra-galactic (EGCR) components are compatible with such a holistic model, provided that the index of the source spectrum be $x \simeq 2.23\pm 0.07$. This is compatible with the best-fit indices for both GCRs and EGCRs, assuming that their source composition is the same, which is indeed the case in a holistic model. It is also compatible with theoretical expectations for particle acceleration at relativistic shocks.
}

%\authorrunning{}
%\titlerunning{}

\maketitle

%**************************************************************************	
%**************************************************************************
\section{Introduction}
\label{sec:Intro}

Despite considerable observational and theoretical progress over the last decades, the origin of cosmic-rays (CRs) remains uncertain at all energies and a subject of intense debate. The very existence of ultra-high energy CRs (UHECRs) challenges the most efficient particle acceleration mechanisms thought to be at work in powerful astrophysical sources, and the CR source model often assumed for low energy particles (involving diffusive shock acceleration in isolated supernova remnants) suffers from a number of persistent problems that shed some doubts about its general validity (e.g. Parizot, et al., 2001, Parizot 2005). The only well-established facts about low-energy CRs are derived from the detailed and joint study of their composition and spectrum between, say, 100 MeV/n and 100 GeV/n, in the light of CR propagation models involving energy losses, in-flight spallation reaction and charged particle diffusion and confinement in a magnetised extended halo (e.g. Strong and Moskalenko, 2001). They can be summarised as follows: phenomenologically, i) the source spectrum is consistent with a power-law, $E^{-x}$, of index $x\simeq 2.2$--2.4 (e.g. Jones, et al., 2001; Ptuskin et al., 2005), ii) the source composition is compatible with the acceleration of the average interstellar medium, somewhat modified by specific selection mechanisms and local enrichment from massive star's ejecta (e.g. Meyer et al., 1997; Ellison, et al., 1997), iii) the diffusion coefficient of the relativistic nuclei increases with their energy-to-charge ratio as $(E/Z)^{\beta}$, where $\beta \simeq 0.3$--0.6 (e.g. Jones, et al., 2001), and iv) the height of the confining halo is relatively large, of the order of 5~kpc above and below the Galactic plane (Moskalenko, et al., 2001). As for the UHECRs, one can typically distinguish between two types of phenomenological scenarios (for astrophysical scenarios with roughly homogeneous extragalactic sources): either pure proton sources with a steep source spectrum in $E^{-2.6}$ or $E^{-2.7}$ or sources with a mixed composition (similar to the low-energy cosmic-rays, say) and a harder source spectrum, in $E^{-2.2}$ or $E^{-2.3}$, typically (see Allard et al., 2005a,b, and refs. therein).

It is customary to distinguish between several regions of the CR spectrum. Some of these distinctions are artificial or related to the history of CR science or to the specific detection technique used in different energy ranges, while others have more phenomenological or theoretical grounds, in relation with observed or expected structures in the spectrum (e.g. due to solar modulation, spatial diffusion, contributions of individual sources, loss of confinement, energy losses...). However, despite the spectral features known as the \emph{knee(s)} and \emph{ankle}, the most striking property of the CR spectrum is its overall regularity and coherence over (at least) 12 orders of magnitude in energy and 32 orders of magnitude in differential flux. This is quite extraordinary and unique for a non-thermal phenomenon. It invites one to consider cosmic rays as a global, unified phenomenon taking place in the universe, and investigate how far it is possible to follow such a road.

However, even within a unified point of view, a distinction cannot be avoided between CRs of Galactic and extragalactic origin. Indeed, low-energy cosmic rays (at least up to TeV energies) are known to be accelerated in our Galaxy, and the lower gamma-ray emissivity from neutral pion decay observed in the large Magellanic cloud (LMC), as compared to the Milky Way, implies that the CR density inside our galaxy is larger than in the LMC (Sreekumar, et al., 1993). This would not be the case if the responsible CRs had an extragalactic origin. At the other end of the spectrum, on the contrary, CRs with an energy larger than, say, $10^{19}$~eV have too large gyroradii to be confined by the Galactic magnetic fields, and their observed (rough) isotropy thus implies an origin outside (of the disk) of the Galaxy. At least two components of CRs are thus observed, one Galactic and one extragalactic, with a transition probably around the so-called \emph{ankle}, at $\sim 3\,10^{18}$~eV (e.g. Allard, et al., 2005a,b) or below (Berezinsky et al., 2004).

However, the fact that the low-energy and high-energy CRs that we observe have a different origin \emph{in space} does not imply that they have a different origin \emph{in nature}. Following a unification perspective looking for as few astrophysical processes as possible behind the CR phenomenon, we can investigate the possibility that all CRs be actually produced by a single type of sources, from thermal energies up to a maximum energy, $E_{max} > 10^{20}$~eV, that still remains to be determined. We refer to such a model as a \emph{holistic model}. Since, as recalled above, the CR sources are quite uncertain at all energies, this opens the possibility that one single model may solve both problems.

Different versions of holistic models have been proposed before, notably with gamma-ray bursts as the source of all CRs, within the cannonball model (Dar and de R{\' u}jula, 2004). In a recent version (Dar, 2004), CRs below and above the knee are accelerated by the same astrophysical objects, but not by the same mechanism nor with the same spectrum. In another approach (Dar and Plaga, 1999), all the observed CRs originate from our galaxy and its halo, with a negligible contribution of other galaxies, even at ultra-high energies. In this \emph{Letter}, we consider a different (strong) version of holistic models, where not only the same source, but also the same acceleration mechanism is operating over the whole spectrum of CRs, with the simplest source spectrum, namely a single power-law, in $E^{-x}$. This is a purely phenomenological study, where the actual CR sources and acceleration mechanism are left unspecified. We simply assume that \emph{some} mechanism produces a CR power-law spectrum in \emph{some} sources located inside galaxies, and investigate the viability of such a generic model in view of the available CR data. We show that the resulting relative normalisation of Galactic and extragalactic components are as observed for a very natural value of the unique free parameter of the model, namely the logarithmic slope of the power-law, $x$.

\section{Relative normalization of Galactic and extragalactic CR components}

In order to compare the observed relative normalisation of the galactic and extragalactic CRs components with the one predicted by the holistic model, we choose a reference energy in each component and compute the corresponding CR density, respectively inside our Galaxy and in the whole universe. We choose $E_{0} = 1$~GeV as the reference value for (low-energy) Galactic cosmic-rays (GCRs), because CR phenomenology (including propagation effects) is well known and constrained by composition measurements around that energy. The CRs injected in the Galaxy are resonantly scattered by magnetic field inhomogeneities and remain confined for a time $\tau_{\mathrm{conf}}(E)$ that depends on their energy. At very high energy, say at $E \ga 5\,10^{18}$~eV, the CR gyroradius is larger than the magnetic field coherence length and even becomes comparable to the size of the confining region, so that the CRs injected inside the Galaxy leak out quickly and diffuse away in the whole universe. Thus, as recalled above, most of the high-energy CRs observed on Earth have an extragalactic origin. We choose $E_{1} = 10^{19}$~eV as the reference value for extragalactic cosmic-rays (EGCRs), because: i) it is high enough for the corresponding CRs to have a diffusion radius (over the energy loss time scale; i.e. a magnetic horizon, see Deligny et al. 2004, Parizot 2004) larger than the distance between typical galaxies, so that their density should be roughly uniform in the universe, and ii) it is low enough for a reliable estimate of the corresponding CR density to be obtained in a simple way, both observationally (since the flux is not yet dominated by statistical uncertainties) and theoretically (since $E_{1}$ is below the GZK regime where stochastic propagation effects and the local source distribution have a noticeable influence on the observed fluxes).

\subsection{GCR density}

Let us assume that CRs are produced in our Galaxy with an average injection rate, in s$^{-1}\,\mathrm{GeV}^{-1}$,
\begin{equation}
Q(E) = Q_{0}\left(\frac{E}{E_{0}}\right)^{-x},
\label{eq:inputSpectrum}
\end{equation}
where the normalisation, $Q_{0}$, relates to the global source power. In practice, it does not matter where and when the CRs are injected in the Galaxy, as long as the granularity of the sources in space and time is small compared to the diffusion radius and confinement time of the CRs in the Galaxy. This amounts to assuming that the observed CRs are not significantly different from what would be observed at another time and somewhere else in the Galaxy, as usually assumed in CR studies.

At low-energy (and in a steady state), the local density of the GCRs is then simply given by the total number of CRs injected during the confinement time, $\tau_{\mathrm{conf}}$, divided by the confinement volume, $V_{\mathrm{conf}}$:
\begin{equation}
n_{\mathrm{G}}(E) \simeq Q_{0}\left(\frac{E}{E_{0}}\right)^{-x}\times \frac{\tau_{\mathrm{conf}}(E)}{V_{\mathrm{conf}}(E)}
\label{eq:nG}
\end{equation}

\subsection{EGCR density}

At high-energy, many (extragalactic) sources contribute to the local density of EGCRs, because of the dilution of the particles over large volumes in the universe. The dilution volume depends on the transport properties in the extragalactic medium. In the absence of magnetic fields, it is merely a sphere of radius $c\tau$, where $\tau$ is the ``lifetime'' of the EGCRs (if shorter than the age of the source), while in a diffusive transport regime with diffusion coefficient $D$, the diffusion radius is given by $\sqrt{6D\tau}$. For an order of magnitude calculation, one may consider that the particles injected at an energy $E$ ``disappear'' when they have lost a substantial fraction of their energy, and thus choose the EGCR lifetime as the energy loss time, $\tau_{\mathrm{loss}}(E)$, defined by the energy drift equation $\mathrm{d}E/\mathrm{d}t = E/\tau_{\mathrm{loss}}(E)$. At the reference energy, $\tau_{\mathrm{loss}}(E_{1}) \simeq 4.4\,10^{9}$~years, which is large enough to ensure an overlap of the dilution spheres around the galaxies.

 %(and are not replenished by higher energy CRs because of the steeply falling injection spectrum)

%Note that above $\simeq 2\,10^{18}$~eV (e.g. Berezinsky ??), the energy losses are dominated by the in-flight production of e$^{+}$e$^{-}$ pairs through interactions with the cosmological radiation field. We shall thus neglect here the energy losses due to universal expansion, as well as the evolution of the source power as a function of time (or redshift), and introduce these effects \emph{a posteriori} as a ``cosmological'' fudge factor, together with uncertainties on the relevant astrophysical quantities.

Then, considering that the EGCRs accumulate uniformly throughout the universe for a time $\tau_{\mathrm{loss}}$, the density of EGCRs is obtained as the number of CRs injected by a galaxy during that time, multiplied by the density of galaxies in \emph{today}'s universe, $n_{gal}$. In practice, galaxies are of different types and one may expect that they inject CRs (roughly) proportionally to their star formation rate. Their contribution to the overall CR density should then be scaled accordingly. Here, we take the Milky Way as a reference and thus use the density of ``Milky Way-like galaxies'' in the universe, estimated from the ratio of the star formation rate density in the local universe, $\simeq 1.4\,10^{-2}\,\mathrm{M}_{\odot}\mathrm{yr}^{-1}\mathrm{Mpc}^{-3}$ (Wyder et al., 2005) and the Galactic star formation rate, $\simeq 3\,\mathrm{M}_{\odot}\mathrm{yr}^{-1}$ (e.g. McKee, 1989): $n_{\mathrm{gal}}\simeq 5\,10^{-3}\,\mathrm{Mpc}^{-3}$. We thus have:
\begin{equation}
n_{\mathrm{EG}}(E) \simeq Q_{0}\left(\frac{E}{E_{0}}\right)^{-x} \times \tau_{\mathrm{loss}}(E)\times n_{\mathrm{gal}}\,,
\label{eq:nEGSimplified}
\end{equation}
to be compared with Eq.~(\ref{eq:nG}).

More rigourously, the EGCR density at energy $E$ is obtained from the sum of all particles injected at earlier times (or redshifts $z$) at a higher energy such that they have lost just enough energy to be observed today at energy $E$:
\begin{equation}
n_{\mathrm{EG}}(E) = n_{\mathrm{gal}}\times \int_{0}^{T}Q(E_{\mathrm{in}}(E,t_{\mathrm{lb}}),t_{\mathrm{lb}})\frac{\mathrm{d}E_{\mathrm{in}}}{\mathrm{d}E}\mathrm{d}t_{\mathrm{lb}},
\label{eq:nEG}
\end{equation}
where $t_{\mathrm{lb}}$ is the look-back time, from today ($t_{\mathrm{lb}} = 0$) back to the onset of EGCR sources ($t_{\mathrm{lb}} = T$), $E_{\mathrm{in}}(E,t_{\mathrm{lb}})$ is the injection energy, at $t_{\mathrm{lb}}$, of EGCRs observed today at energy $E$, and $Q(E,t) = Q_{0}(E/E_{0})^{-x}\times f(t)$,  allowing for a possible evolution of the source power ($f(0) = 1$). The injection energy is computed by solving the energy loss equation $\mathrm{d}E/\mathrm{d}t = \dot{E}(E,t)$, where the latter function includes both expansion losses,
\begin{equation}
\frac{\mathrm{d}E}{\mathrm{d}t}\bigg |_{\mathrm{exp}} = \frac{E}{1+z} \times \frac{\mathrm{d}t}{\mathrm{d}z},
\label{eq:dEDtExpansion}
\end{equation}
and losses due to the EGCR interaction with the cosmological background radiation (e.g. Berezinsky and Grigorieva, 1988):
\begin{equation}
\frac{\mathrm{d}E}{\mathrm{d}t}\bigg |_{\mathrm{CMB}} = E\,(1+z)^{3}\,K\int_{0}^{\infty}\hspace{-7pt}\langle\sigma\kappa\rangle[xE(1+z)] \,x\ln(1 - e^{-\beta x})\mathrm{d}x,
\label{eq:dEDtCMB}
\end{equation}
where $\langle\sigma\kappa\rangle$ is the product of the interaction cross section (for e$^{+}$-e$^{-}$  and pion production) by the corresponding inelasticity, the constants $K$ and $\beta$ are given by $K = (k_{\mathrm{B}}T_{0}m_{\mathrm{p}}^{2}c^{4})/(2\pi\hbar^{3}c^{2}E)$ and $\beta = m_{\mathrm{p}}c^{2}/2k_{\mathrm{B}}T_{0}$, and $T_{0} = 2.73$~K.

\begin{figure}[t]
\includegraphics[width=\linewidth]{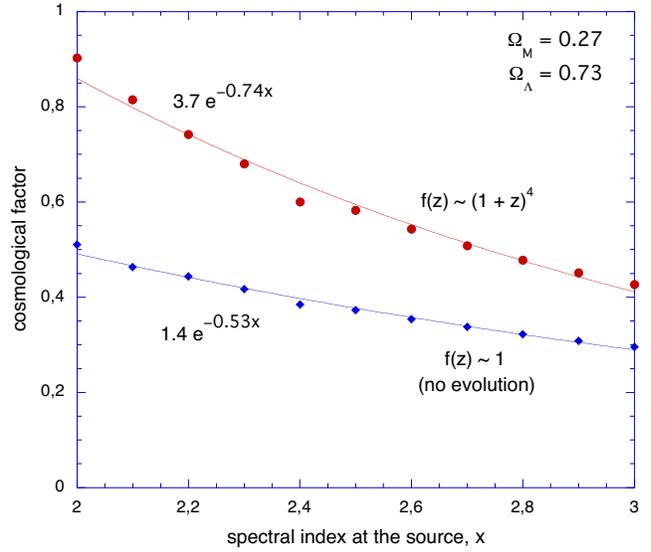}
\caption{Cosmological factor, $\mathcal{I}(E_{1},x)$, defined in Eq.~(\ref{eq:ICosmo}), as a function of the CR source spectrum index, $x$, for $E_{1} = 10$~EeV and two different choices of the source power evolution, $f(z)$.}
\label{fig:ICosmo}
\end{figure}

Equation~(\ref{eq:nEG}) can be rewritten as
\begin{equation}
n_{\mathrm{EG}}(E) \simeq Q_{0}\left(\frac{E}{E_{0}}\right)^{-x}\times \tau_{\mathrm{loss}}(E)\times n_{\mathrm{gal}}\times \mathcal{I}(E,x),
\label{eq:nEGBis}
\end{equation}
where $\tau_{\mathrm{loss}} = E/\dot{E}(E,0)$ is computed at $z = 0$ and the integral $\mathcal{I}(E,x)$ gathers all the modifications to the simplified Eq.~(\ref{eq:nEGSimplified}) due to the cosmological evolution of the sources and the time-dependent energy losses:
\begin{equation}
\mathcal{I}(E,x) \equiv \frac{\dot{E}(E,0)}{E}\int_{0}^{\infty} \left(\frac{E_{\mathrm{in}}}{E}\right)^{-x}\frac{\mathrm{d}E_{\mathrm{in}}}{\mathrm{d}E}f(z)\frac{\mathrm{d}t}{\mathrm{d}z}\,\mathrm{d}z.
\label{eq:ICosmo}
\end{equation}
Note that $\mathcal{I}(E,x)$ would be equal to $1/(x -1)$ if $\dot{E}(E,t)$ were constant and the sources did not evolve, i.e. $f(z) = 1$. If $H_{0}$ is the current Hubble ``constant'', $\Omega_{\mathrm{M}}$ is the usual normalised matter density, $\Omega_{\mathrm{\Lambda}}$ is the reduced cosmological constant and $\Omega_{\mathrm{k}} = 1 -  \Omega_{\mathrm{M}} - \Omega_{\mathrm{\Lambda}}$ (the ``curvature density''), we have:
\begin{equation}
\frac{\mathrm{d}t}{\mathrm{d}z} = \frac{1}{H_{0}(1+z)^{2}}\left[\Omega_{\mathrm{M}}(1+z) + \Omega_{\mathrm{k}} + \frac{\Omega_{\Lambda}}{(1+z)^{2}}\right]^{-1/2}.
\label{eq:dTDz}
\end{equation}
This allows us to compute $\mathcal{I}(E,x)$, as shown on Fig.~\ref{fig:ICosmo} (we assume $\Omega_{\mathrm{M}} = 0.27$, $\Omega_{\mathrm{\Lambda}} = 0.73$ and $H_{0} \simeq 75$ km/s/Mpc). It is convenient here to approximate $\mathcal{I}(E,x)$ by an exponential function of $x$. To within a few percent in the range of $x$ of interest, we found $\mathcal{I}(E_{1},x) \simeq 1.4\,e^{-0.53 x}$ in the case when $f(z) = 1$ (no source evolution) and $\mathcal{I}(E_{1},x) \simeq 3.7\,e^{-0.74 x}$ with a source evolution in $f(z) = (1+z)^{4}$ (by analogy with the star formation rate).

\subsection{GCR/EGCR normalisation}

The predicted value of $n_{\mathrm{G}}(E_{0})/n_{\mathrm{EG}}(E_{1})$ follows straightforwardly from Eqs.~(\ref{eq:nG}) and~(\ref{eq:nEGBis}):
\begin{equation}
\frac{n_{\mathrm{G}}(E_{0})}{n_{\mathrm{EG}}(E_{1})} \simeq \left(\frac{E_{1}}{E_{0}}\right)^{x} \times \,\frac{\tau_{\mathrm{conf}}(E_{0})}{\tau_{\mathrm{loss}}(E_{1})}\times \,\frac{1}{n_{\mathrm{gal}}V_{\mathrm{conf}}(E_{0})\mathcal{I}(E_{1},x)}
\label{eq:nGOverNEG}
\end{equation}
Identifying the left hand side with the measured CR flux ratio and using writing $\mathcal{I}(E_{1},x)\simeq Ae^{-B x}$ (see above), one can easily solve the above equation to determine the value of the only free parameter in the model, namely the logarithmic index of the (holistic) injection spectrum, $x$:
\begin{equation}
x = \ln\left[A\, n_{\mathrm{gal}}V_{\mathrm{conf}}(E_{0})\frac{\tau_{\mathrm{loss}}(E_{1})}{\tau_{\mathrm{conf}}(E_{0})} \,\frac{\Phi_{\mathrm{CR}}(E_{0})}{\Phi_{\mathrm{CR}}(E_{1})}\right] \Big/ \ln\left[\frac{E_{1}}{E_{0}}\, e^{B}\right].
\label{eq:alpha}
\end{equation}
%\begin{equation}
%x = \ln\left[3.7\frac{n_{\mathrm{gal}}V_{\mathrm{conf}}(E_{0})\tau_{\mathrm{loss}}(E_{1})}{\tau_{\mathrm{conf}}(E_{0})} \,\frac{\Phi_{\mathrm{CR}}(E_{0})}{\Phi_{\mathrm{CR}}(E_{1})}\right] \Big/ \ln\left[\frac{E_{1}}{E_{0}}\times e^{0.74}\right].
%\label{eq:alpha}
%\end{equation}

The required value of $x$ is thus given here as a function of parameters that can be measured or derived directly from astrophysical data. Numerically, we gave above the value of $\tau_{\mathrm{loss}}(E_{1})$ and an estimate of $n_{\mathrm{gal}}$. The confinement time at 1~GeV if roughly $\tau_{\mathrm{conf}}(E_{0}) \simeq 2.4\,10^{7}$~years (Connel, 1998). As recalled above, recent propagation studies favour halo heights of the order of 5~kpc (above and below the Galactic plane). Assuming a CR disk radius of 20~kpc, one may thus estimate the confinement volume at 1~GeV as $V_{\mathrm{conf}}(E_{0}) \simeq \pi\times (20\,\mathrm{kpc})^2\times 10\,\mathrm{kpc}\simeq 1.3\,10^{-5}\,\mathrm{Mpc}^3$. For the measured CR flux at $E_{0}$ we take the differential flux deconvoluted from solar modulation (Webber, 1998): $\Phi_{\mathrm{CR}}(1\,\mathrm{GeV})\simeq 0.5\,\mathrm{cm}^{-2}\mathrm{sr}^{-1}\mathrm{s}^{-1}\mathrm{GeV}^{-1}$, and at $E_{1}$ we take an average of AGASA (Teshima, et al., 2004) and HiRes (Zech, et al., 2004) values: $\Phi_{\mathrm{CR}}(10^{19}\,\mathrm{eV})\simeq 2\,10^{-28}\,\mathrm{cm}^{-2}\mathrm{sr}^{-1}\mathrm{s}^{-1}\mathrm{GeV}^{-1}$. Therefore, $\Phi_{\mathrm{CR}}(E_{0})/\Phi_{\mathrm{CR}}(E_{1}) \simeq 2.5\,10^{27}$. Inserting these values into Eq.~(\ref{eq:alpha}), and assuming either $f(z) = (1 + z)^{4}$, so that $A = 3.7$ and $B = 0.74$, or $f(z) = 1$, so that $A = 1.4$ and $B = 0.53$, we find:
\begin{equation}
\begin{split}
&x \simeq 2.23 \pm 0.07,\,\mathrm{for}\,f(z) = (1 + z)^{4},\\
&x \simeq 2.21 \pm 0.07,\,\mathrm{for}\,f(z) = 1,
\end{split}
\label{eq:alphaNum}
\end{equation}
%\begin{equation}
%x \simeq 2.23 \pm 0.07,
%\label{eq:alphaNum}
%\end{equation}
where the ``error bars'' account for an indicative uncertainty of a factor of 5 on the quantity $n_{\mathrm{gal}}V_{\mathrm{conf}}(E_{0})(\tau_{\mathrm{loss}}(E_{1}) / \tau_{\mathrm{conf}}(E_{0}))$, as well as a possibly different source evolution function. %(all other parameters are known with a much better precision).

\section{Discussion}

%[NB: as we have said that $x$ is the only free parameter. However, there are implicit parameters, such as $Q_{0}$, $\beta$, etc. However, these parameters are common to any model of CRs, and...]

The result shown in Eq.~(\ref{eq:alphaNum}) reads as follows: holistic CR source models (where CRs of all energies are produced by the same sources with a single power-law spectrum) are indeed possible, and the corresponding spectral index is determined unequivocally from measured parameters: $x \simeq 2.2$--2.3. This is remarkable in several respects. Not only could a sensible value of $x$ be found, but the solution is particularly interesting from the phenomenological and theoretical points of view. This slope is indeed in keeping with common expectations, and in remarkable agreement with relativistic shock acceleration predictions (e.g. Kirk et al., 2000). It is also right in the range allowed for low-energy CRs, as derived from detailed studies of the secondary-to-primary composition ratios (cf. Sect.~\ref{sec:Intro}).

The above solution is also particularly interesting in relation with EGCRs. While the best fit of the highest-energy data is usually obtained with a source spectrum in $E^{-2.6}$ (De Marco, et al., 2003), this result only holds for pure proton sources. Most (if not all) astrophysical sources of CRs would however accelerate heavier nuclei just as well, since electromagnetic processes only depend on the charged particle rigidity. Quite remarkably, it is found that when assuming a similar composition for EGCRs and GCRs (as would of course naturally be the case in a holistic model), the high-energy data are best reproduced by a power-law source spectrum in $E^{-2.2}$ or $E^{-2.3}$, depending on source evolution (Allard et al., 2005a,b). Thus, the source spectrum that makes holistic models viable is precisely the one that is consistently favoured at low energy (by GCR studies) and, independently, at high energy (by EGCR studies).

Even though it is simple, the above analysis is quite robust. The most uncertain parameters are the CR confinement volume at 1~GeV and the density of Milky-Way-equivalent galaxies in today's universe. However, even large variations of these parameters cannot change the value of $x$ substantially, as a result of the huge lever arm between low-energy GCRs and high-energy EGCRs. It should also be noted that the determination of $x$ does not depend on any assumption concerning the energy dependence of the CR confinement time. We simply used the value derived from the CR data at $E_{0} = 1$~GeV. Interestingly, however, it then leads to a determination of the confinement time that would allow one to reproduce the observed GCR spectrum, in $\sim E^{-2.71}$: $\tau_{\mathrm{conf}}(E) \propto E^{\delta}$, with $0.41 \la \delta \la 0.55$, which is also the range expected from CR diffusion theory. More generally, all the phenomenological studies concerning GCRs and EGCRs turn out to be equally valid within the holistic models, because the required spectral index is the one that was derived (or needed) at both low and high energies anyway.

In conclusion, we found that a holistic CR source model with a single power-law spectrum in $E^{-x}$ and $x \simeq 2.23\pm 0.07$ could account for the CRs at all energies, in keeping with all known results about CR phenomenology, and in agreement with the main theoretical results for CR acceleration and transport. However, it should be clear that we did not propose a well-defined physical or astrophysical model. We simply made a general phenomenological remark that may motivate or encourage further studies of the origin of CRs considered globally, as a general phenomenon possibly involving a single process at work over their whole energy range.

Finally, one can propose another reading of the result obtained here. Reversing the point of view, one may state that, whatever the sources of GCRs, \emph{if} their spectrum actually extends up to $\sim 10^{20}$~eV, then the contribution of the corresponding high energy particles escaping from all galaxies is just enough to explain the EGCR fluxes and properties (i.e. no other sources are needed), \emph{provided that} the source spectral index is $\sim 2.2$--2.3. Likewise -- and perhaps more strikingly--, even though we do not know what the EGCR sources are, it is very natural to expect that they do not produce \emph{only} UHECRs, and that the latter are just the high-energy end of a continuous spectrum starting at thermal energies in the sources. Now, according to our result, \emph{if} the EGCR source spectrum extends down to non-relativistic energies indeed, then the contribution of the corresponding low-energy CRs confined within the parent galaxies is just enough to explain the GCR flux and properties as well (i.e. no other sources are needed), \emph{provided that} the source spectral index is $\sim 2.2$--2.3, which it is inferred to be anyway, from (mixed-composition) EGCR studies.

\begin{acknowledgements} EP thanks Arnon Dar and Paul Sommers for interesting discussions about the model.
\end{acknowledgements}

\end{document}